# Rural Access Index: A global study

Quan Sun [a], Wanjing Li [a,b], Qi Zhou [*]

[a] School of Geography and Information Engineering, China University of Geosciences – Sunquan@cug.edu.cn,
[b] School of Geography and Information Engineering, China University of Geosciences

[*] Corresponding author

**Abstract**: The Rural Access Index (RAI), one of the UN Sustainable Development Goal indicators (SDG 9.1.1), represents the proportion of the rural population residing within 2 km of all-season roads. It reflects the accessibility of rural residents to transportation services and could provide guidance for the improvement of road infrastructure. The primary deficiencies in assessing the RAI include the limited studying area, its incomplete meaning and the absence of correlation analysis with other influencing factors. To address these issues, this study proposes the "Not-served Rural Population (NSRP)" as a complementary indicator to RAI. Utilizing multi-source open data, we analysed the spatial patterns of RAI and NSRP indicators for 203 countries and then explored the correlation between these 2 indicators and other 10 relevant factors. The main findings are as follows: 1) North America, Europe, and Oceania exhibit relatively high RAI values (>80%) and low NSRP values (<1 million). In contrast, African regions have relatively low RAI values (<40%) and high NSRP values (>5 million). There is a negative correlation between RAI and NSRP. 2) There is spatial autocorrelation and significant imbalances in the distribution of these two indicators. 3) The RAI exhibit a positive correlation with the factors showing levels of the development of countries such as GDP, education, indicating that improving the road infrastructure could reduce the poverty rates and enhance access to education. And in contrast with RAI, NSRP exhibit the completely negative correlations with these factors.

## 1. Introduction

The Rural Access Index (RAI) is defined as "the proportion of the rural population residing within 2 km of all-season roads ". It was included as one of the indicators for achieving United Nations Sustainable Development Goal 9 - Industry, Innovation, and Infrastructure in 2017. (IAEG-SDGs; UN General Assembly 2017). The RAI reflects the accessibility of rural people to road services(World Bank 2003a; World Bank 2003b) making it a crucial factor for policymakers and managers to consider when analysing road accessibility. Evaluating RAI could also provide supports for the improvement of the road infrastructure. This, in turn, ensures that rural populations can fully benefit from socioeconomic development initiatives.

Currently, numerous scholars have utilized different means of GIS to calculate and analyse the RAI in various countries. For instance, Workman and McPherson (2019) calculated RAI values for Ghana, Malawi, Myanmar, and Nepal and then proposed some suggestions for improving RAI calculations. Ilie et al. (2019) described their model for calculating RAI and applied it to Tabora and Tanzania. Wahyuni et al. (2022) calculated the RAI at the county level in Indonesia from 2014 to 2020 and evaluated the impact of government-supported road infrastructure projects. They found that during the period of the project, not only did the RAI improve significantly, but the regional inequality also decreased. An analysis of the spatial pattern of RAI conducted by Li et al. (2021) showed that the RAI values in northern African countries were relatively high.

Furthermore, some scholars have also explored factors influencing road accessibility, including GDP, DEM, agriculture, medical services, etc. Wondemu and Weiss (2012) illustrated the relationship between the road accessibility and income in the case of rural Ethiopia. They found that improving the rural accessibility was a vital approach to increase the income. Zhou et al. (2021) found that road construction could significantly promote the sustainable development of agriculture through a study of 31 provinces in China. Nidup (2016) analysed the relationship between road accessibility and poverty rates in Bhutan. The findings indicated that the better rural road accessibility contributed to the higher wealth accumulation and lower poverty rates among poor households. Kwigizile et al. (2011) analysed four villages in Tanzania and discovered that the impoverished people with high accessibility to rural roads had more opportunities to escape poverty and a higher awareness of poverty alleviation.

However, the existing studies about the RAI have predominantly concentrated on less developed countries and regions, particularly in Africa. These studies have not only been limited in scope, but also primarily focusing on calculating the values of RAI within their studying areas. (Vincent S 2018; Piloyan et al. 2018; Mikou et al. 2019) There is a lack of analysis on the global scale and insufficient exploration of the correlation between the spatial pattern of RAI and its influencing factors.

In recent years, there has been a significant increase in the availability of diverse open datasets, which are characterized by their free access, global coverage, and rich geographical information. Consequently, they are recognized as vital resources for assessing the United Nations' Sustainable Development Goals (Guo H et al. 2016; Bennett J 2010) The open datasets also play a crucial role in the calculation of RAI. Recognizing these, this study proposed" Not-served Rural Population (NSRP)" to complement RAI. Based on the open datasets, the study aims to achieve 3 primary objectives: 1) Propose a new method for calculating RAI based on multi-source open data. 2) Reveal the spatial patterns of the RAI on the global scale. 3) Quantitatively assess the relationship between RAI and other pertinent factors such as socio-economic, employment, poverty levels, education, and more.

The main findings of this study can assist policymakers to make reasonable policies for enhancing rural road accessibility, so as to promote socio-economic development and bridge regional disparities.

This paper is structured as follows: Section 2 describes the data from different sources and some details. Section 3 introduces the specific methods and the indicators used to analyse in this study. Section 4 presents the results of different analysis. Section 5 conclude the findings of the study.

## 2. Data

The data used in this study include: population data, urban built-up area data, road data, administrative boundary data, and socio-economic data. Since national road data were hard to obtain directly from countries worldwide, this study calculated the RAI and NSRP indicators for different countries and regions based on the road data from OpenStreetMap. The details of the data are as follows:

(1) Road data

OpenStreetMap (OSM), which provides various geospatial data including roads, water, land cover, etc., is a typical example of voluntary geographic information. It is a free, open-source, and editable online map that volunteers worldwide create and update collaboratively. (Haklay M 2008) In this study, global road data for January 2020 were obtained from the OpenStreetMap (OSM) official website (https://www.openstreetmap.org/).

(2) Administrative boundary data

The global administrative boundary data came from the Database of Global Administrative Areas (https://gadm.org/data.html). This dataset is a highly accurate global administrative boundary dataset that contains different levels of administrative boundaries for most countries and regions worldwide.

(3) Socio-economic data

Global socio-economic and related data came from the World Bank (https://data.worldbank.org/indicator). The data years were all between 2018 and 2020.

(4) Population data

Developed by the WorldPop team at the University of Southampton, the WorldPop is based on the data including roads, elevation, land cover, night-time lights and GPW. (Steven et al. 2015) The WorldPop dataset includes 2 different resolutions, 1km and 100m, and the product with higher resolution(100m) was selected for this study.

(5) Urban built-up area data

The Urban built-up area data with 1km resolution utilized for extracting the rural area in this study comes from the Global Dataset of Annual Urban Extents produced by Zhao et al (2022). This dataset, which is produced based on the night-time lights, has the advantages of long time series, wide coverage and high reliability.

## 3. Methodology

This study aims to analyse the spatial patterns and associated factors of global RAI and NSRP. We seek to address scientific inquiries about the spatial heterogeneity of the global RAI and explore its relationship with other influencing factors.

In pursuit of these objectives, we firstly identified OpenStreetMap (OSM) tags that ensure automobile access, including "trunk", "primary", "secondary", "tertiary", "unclassified", "residential", "living_street", "road", "trunk_link", "primary_link", "secondary_link" and "tertiary_link". Employing the prescribed methods outlined in the Section 3.1, this study computed the values and of RAI and NSRP for 203 countries and regions across the globe and visualized the distributions of these 2 indicators.

Subsequently, based on the autocorrelation analysis (Moran's I) mentioned in Section 3.2 and the inequality analysis (Gini coefficient) mentioned in Section 3.3, the characteristics of spatial patterns and the inequality of these indicators were analysed.

Lastly, considering variables from 5 different dimensions, including population, economy, employment, poverty and education, the relationship between the global RAI and NSRP indicators and other influencing factors was explained quantitatively through correlation analysis.

The subsequent sections provide specific introductions to these methods.

### 3.1 Calculation of RAI and NSRP

(1) Rural Access Index

The Rural Access Index (RAI) represents the proportion of people residing within 2 km of all-season roads. Its formula is as follows:

$$RAI = \frac{POP_{served}}{POP_{rural}} \times 100\% \qquad (1)$$

Where, $RAI$ denotes RAI. $POP_{rural}$ is the total number of rural people. $POP_{served}$ is the number of rural people residing within 2 km of all-season roads. The RAI ranges from 0% to 100%. When the RAI equals 100%, it indicates that all the rural people are living within 2 km of all-season roads.

(2) Not-served Rural Population

Since the RAI is just a percentage which cannot measure the specific number of rural people facing with the challenges of accessing road services, this study proposed NSRP indicator in order to calculate the number of these rural people. The formula is as follows:

$$NSRP = POP_{rural} - POP_{served} \qquad (2)$$

Where *NSRP* indicates the number of rural populations who have difficulties accessing road services. The meanings of *POPrural* and *POPserved* are the same as those in the formula of calculating RAI.

The RAI and NSRP essentially demonstrate the difficulties faced by rural populations in accessing road services through analysing the distributions of rural people and all-season roads. Therefore, there are two primary challenges needed to be considered for the automatic calculation of these two indicators: 1) How to extract the distribution of rural populations and all-season roads? 2) How to get RAI and NSRP through the relationships between the datas used in the study.

To address these two problems, this study proposed an automatic method of calculating RAI and NSRP based on the multi-source open data. To extract the rural population distribution firstly, we overlayed the administrative boundary data and the urban built-up data in order to obtain the non-urban built-up areas as the rural areas, which were used for getting the total rural populations by overlaying the population data. Secondly, this study filtered the spatial database to extract the distribution of all-season access roads in the rural area. Specifically, the highways and the elevated roads were excluded since they may be impassable due to the impact of severe weathers sometimes. To explore the relationship between the distribution of rural populations and all-season roads, this study calculated the nearest distance from the center of the population grid to the all-season roads (through proximity analysis) and then revised this distance by DEM. The formula for revising the distance is as follow:

$$D' = \sqrt{D^2 + (E_{pop} - E_{road})^2} \qquad (3)$$

Where *D'* represents the distance after being revised. *D* is the distance from the center of the grid of the population to the nearest vertical point of the all-season road. $E_{pop}$ and $E_{road}$ are the elevations of the center of the population grid and the nearest vertical point respectively.

Taking a single county unit as an example, Figure 1 illustrates the process of the calculation of RAI and NSRP indicators

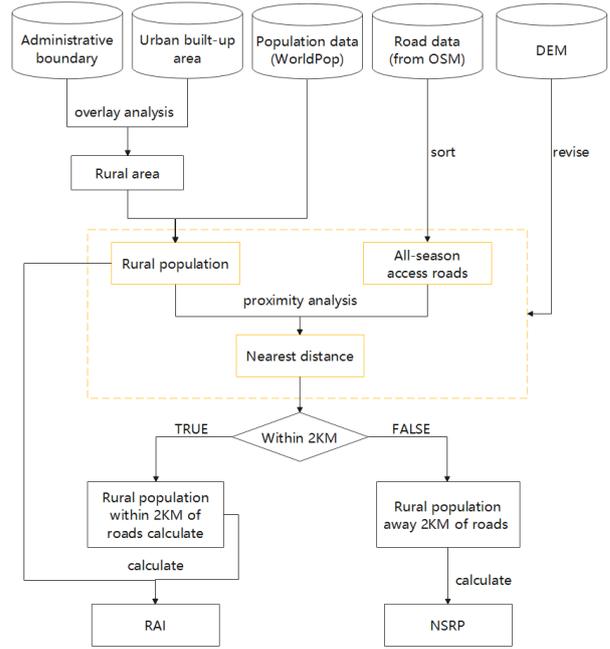

Figure 1 The process of calculating RAI and NSRP

### 3.2 Spatial Pattern Analysis

This study calculated the values of global Moran's I and Gini coefficients in order to delve the spatial patterns of these two indicators from two different dimensions, namely the spatial autocorrelations and their inequality in space.

#### 3.2.1 *Spatial Autocorrelation Analysis*

The spatial autocorrelation analysis is a significant way to examine if the objects of the study exhibit clustering, dispersion, or randomness in their spatial distributions (Mitchel A 2005; Anselin L 1995). The Moran's I, which is calculated to represent the autocorrelation pattern, is divided into 2 categories, namely the Global Moran's I(calculating on the global scale) and Local Moran's I(calculating on a small scale). This study employed Global Moran's I to explore which category of the autocorrelation the RAI and NSRP present in space.

The formulas for calculating global Moran's I are as follows:

$$I = \frac{\sum_{i=1}^{n}\sum_{j=1}^{n} w_{ij} s_{ij}}{\sigma^2 \sum_{i=1}^{n}\sum_{i=1}^{n} w_{ij}} = \frac{\sum_{i=1}^{n}\sum_{j=1}^{n}(x_i - \bar{x})(x_j - \bar{x})}{\sigma^2 \sum_{i=1}^{n}\sum_{i=1}^{n} w_{ij}} \qquad (3)$$

$$\sigma^2 = \sum_{i=1}^{n} \frac{(x_i - \bar{x})^2}{n} \qquad (4)$$

$$\bar{x} = \frac{1}{n}\sum_{i=1}^{n} x_i \qquad (5)$$

Where *n* is the number of the objects in the study area, i.e., the number of the countries and regions worldwide. $\sigma^2$ is the overall variance of the specific attribute characteristics '*x*' of the study objects, i.e., the overall variance of the RAI or NSRP indicators. $\bar{x}$ is the mean value of the specific attribute characteristics '$x$', i.e., the mean value of RAI or NSRP indicators. $x_i$ and $x_j$ represent the observed values of the specific attribute feature '$x$', i.e., the RAI or NSRP indicators at spatial locations $(i,j)$. $W_{ij}$ is the spatial weight. The specific introductions of autocorrelation analysis could refer to Local Indicators of Spatial Association (Anselin L 1995).

### 3.2.2 *Inequality Analysis*

The Gini coefficient, which is derived from the Lorenz curve, has been widely utilized to quantify the level of fairness in the distribution. The Gini coefficient ranges from 0 to 1, with a higher value indicating the larger inequality of the distribution. (Dorfman R 1979; Milanovic B 1997) In this study, the Gini coefficients of RAI and NSRP were calculated respectively so as to explain the inequality of the distributions of these two indicators in space, which could reveal the regional development disparities to some extent.

The formula for calculating Gini coefficient is as follow:

$$G = 1 - \frac{S_1}{5000} \quad (7)$$

Where $G$ is the Gini coefficient and $S_1$ is calculated according to equation (7):

$$S_1 = \frac{1}{2}\sum_{i=1}^{n}(Y_i + Y_{i+1})h_{i+1} \quad (8)$$

Where $Y_i$ is the cumulative proportion of resources, i.e. the value of RAI and NSRP indicators. $h_{i+1}$ represents the total population proportion of the corresponding country. $i$ is the set sequence of those countries or regions.

The Gini coefficient ranges from 0 to 1 and the meaning is demonstrated in Table 2.

Table 2 Explanation of Gini coefficient

| Value range | Description |
|---|---|
| <0.2 | Minimal inequality in RAI/NSRP between countries worldwide, highly equal |
| 0.2-0.29 | Small inequality in RAI/NSRP between countries worldwide, relatively equal |
| 0.3-0.39 | Moderate inequality in RAI/NSRP between countries worldwide, relatively reasonable |
| 0.4-0.59 | Large inequality in RAI/NSRP between countries worldwide |
| >0.6 | Extremely large inequality in RAI/NSRP between countries worldwide |

## 3.3 Correlation Analysis

### 3.3.1 *Selection of variables.*

This study chose relevant factors from five dimensions: population, economy, employment, poverty, and education, aiming to quantitatively examine the relationship between global RAI and NSRP indicators and these five categories of factors.

The specific indices are presented in Table 3, comprising a total of 10 indicators that are available for download from the official website of the World Bank. All indicators have a statistical significance, with each having a count greater than 70.

Table 3 Selections of Variables

| Dimension | Specific indicators | Number of indicators |
|---|---|---|
| Population | Rural Population | 193 |
| | The population people living in rural areas | 193 |
| Economy | GDP per capita | 185 |
| | GDP per employed person | 171 |
| Employment | Employment rate | 181 |
| | Total unemployment rate | 181 |
| Poverty | Proportion of impoverished population | 70 |
| | Gini coefficient (Income level) | 85 |
| Education | Pre-school education rate | 136 |
| | Adult literacy rate | 75 |

1) Population: The Rural Access Index (RAI) denotes the percentage of the rural population residing within a two-km distance of all-season access roads. As its definition suggests, RAI is associated with population size, specifically the rural population. (Roberts et al. 2006; World Bank 2016a; World Bank 2016b) Thus, the rural population and the proportion of rural population were included.

2) Economy: Road accessibility is connected to the level of economic development. (Khandker et al. 2011) Therefore, two key indicators, namely GDP per capita and GDP per capita of the employed population, were included in the economic dimension of this study.

3) Employment: Previous studies demonstrate the influence of rural road accessibility on employment (Fiorini et al. 2022; Gibson et al. 2003). Consequently, this study selected the employment rate and the proportion of the total unemployed population as specific indicators for this dimension.

4) Poverty: Road accessibility is able to improve the levels of local poverty (CIESIN S 2015; Steven et al. 2015; Iimi et al. 2016; Adukia et al. 2020). This study considered two primary indicators in the poverty dimension: the proportion of the impoverished population and the Gini coefficient (reflecting income level).

5) Education: Road accessibility can also contribute to the development of local education (United Nations 2017). Therefore, this study included the preschool education rate and the literacy rate of adults as specific indicators for the education dimension.

### 3.3.2 *Pearson Correlation analysis.*

Pearson correlation analysis is an effective way to measure the correlation between the studying objectives.

For the quantitative assessment of the relationships between the RAI and NSRP indicators, along with other influencing variables (such as socioeconomic factors, poverty, etc.), this study employed Pearson correlation coefficients to quantitatively explore the potential correlations. Its formula is as follow:

$$Pearson(x, y) = \frac{\text{cov}(x, y)}{\sigma_x \sigma_y} = \frac{\sum xy - \frac{\sum x \sum y}{N}}{\sqrt{(\sum x^2 - \frac{(\sum x)^2}{N})(\sum y^2 - \frac{(\sum y)^2}{N})}} \quad (9)$$

Where *Pearson(x,y)* represents the Pearson correlation coefficients. *X* and *Y* are the values of the RAI and NSRP of the studying unit or county, respectively. *Cov(x,y)* denotes the covariance between *x* and *y*. $\sigma_x$ and $\sigma_y$ are the standard deviation of variables *x* and *y* respectively and *N* denotes the sample size.

## 4. Results and Analysis

### 4.1 Spatial pattern analysis

Applying the calculation methods elucidated in Section 3.1, the study got the average RAI value to be 68.26% and cumulative NSRP amounts to 1.173 billion people across 203 countries and regions worldwide. Figure 2(a) and Figure 2(b) illustrate the spatial distributions of the RAI and NSRP across 203 countries worldwide respectively.

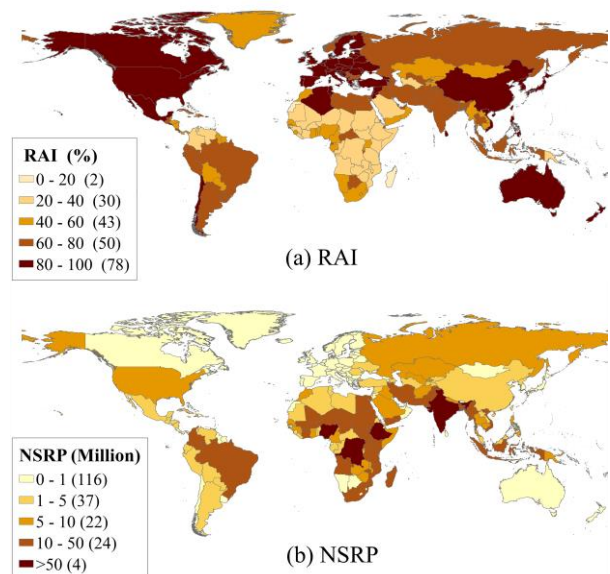

Figure 2: Global Distributions of RAI and NSRP

From the perspective of spatial pattern, countries with high RAI values are mainly located in North America (e.g., the United States and Canada), Europe (e.g., France and Italy), Eastern Asia (e.g., China and Japan), and Oceania (e.g., Australia and New Zealand). Notably, North America and Oceania exhibit the highest RAI values, both exceeding 80%. However, countries in the northern regions of South America (e.g., Colombia and Venezuela) and Africa (e.g., Mali and Niger) demonstrate relatively low RAI values. Most African countries have the lower RAI (all within 20% to 40% range) than countries or regions in other continents. In particular, Western Sahara and Madagascar, both located in Africa, possess the lowest RAI values globally, standing at 18.93% and 16.13%, respectively.

Table 4 shows the spatial indicators of the RAI and NSRP at the global level.

Table 4 The results of the spatial autocorrelation

| Indicators | RAI | NSRP |
|---|---|---|
| Global Moran's I | 0.381 | 0.317 |
| Expectation Index | -0.004 | -0.004 |
| Variance | 0.003 | 0.002 |
| z-score | 7.093 | 8.167 |
| p value | 0.000 | 0.000 |
| Gini coefficient | 0.814 | 0.557 |

According to the findings presented in Table 4, the Moran's I value of global RAI is 0.381, with a z-score of 7.093 and a p-value of 0.000. Applying the null hypothesis principle, when the p-value is below 0.01 and the z-score exceeds 2.58, it can be considered that the global RAI distribution is non-random. Since the Moran's I value is greater than 0, it signifies that neighboring countries tend to have similar RAI values. For instance, countries or regions with higher RAI values (e.g., France and Italy) tend to be surrounded by other countries or regions also having higher RAI values, and vice versa for those with lower RAI values.

On the other hand, countries with relatively low NSRP values are primarily located in Northern North America (e.g., Canada), Europe (e.g., France), and Oceania (e.g., Australia). It is worth noting that all countries in Oceania have an NSRP of less than 1 million. However, Africa (e.g., Ethiopia) and Southwest Asia (e.g., India, Bangladesh and Myanmar) exhibit relatively high NSRP values. Specifically, all countries in Africa have NSRPs greater than 5 million, with 18 countries (e.g., Sudan and Uganda) exceeding an NSRP of 10 million. The two countries with the highest NSRPs globally are India and Nigeria, whose values are 230 million and 75 million respectively. The global Moran's I value of NSRP is 0.317 with a z-score of 8.167 and a p-value of 0.000, which also suggests the presence of clustering among countries.

The Gini coefficients of RAI and NSRP, both exceeding 0.4, indicate relatively strong inequality of the RAI on the global scale. As is aforementioned above, it is manifested that there is a huge difference of RAI values between western countries (e.g., Europe and North America) and African countries. The large difference of RAI between countries also reflects the huge disparities of the regional development to some extent.

### 4.2 Correlation Analysis

Table 5 presents the correlation among RAI, NSRP, and other explanatory variables.

Table 5 Pearson Correlation table

| Factor Name | RAI | NSRP |
|---|---|---|
| Rural Population | 0.013 | 0.823** |
| Proportion of Rural Population | -0.462** | 0.234** |
| Per Capita GDP | 0.448** | -0.142* |
| Per Capita GDP of Employed Population | 0.500** | -0.154* |
| Employment Rate | -0.200** | 0.094 |
| Total unemployment rate | -0.061 | -0.034 |
| Proportion of Impoverished Population | -0.349** | 0.241** |
| Gini Coefficient (Income Level) | 0.106 | 0.093 |
| Pre-school Education Rate | 0.445** | -0.121 |
| Adult Literacy Rate | -0.015 | 0.113 |

Note: * $p<0.05$, ** $p<0.01$.

As shown in Table 5, RAI exhibits a significant positive correlation with per capita GDP, per capita GDP of the employed population, and the preschool education rate. And RAI shows a significant negative correlation with the proportion of rural population, employment rate, and the proportion of impoverished population. In contrast with RAI, the NSRP shows almost opposite correlations with these indicators above.

It is evident that countries with high per capita GDP (Figure 5(c)) high per capita GDP of the employed population (Figure 5(d)) and high pre-school education rates (Figure 5(g)) are primarily located in Europe (e.g., Norway, France and Italy), North America (e.g., the United States) and Oceania (e.g., Australia). Conversely, countries with low these values are mainly situated in Africa (e.g., Burundi, Mali and Niger) and some parts of Asia (e.g., Iran and Bangladesh) and South America (e.g., Brazil, Ecuador). On the other hand, it illustrates that countries in most African regions (e.g., Niger, Ethiopia and Chad) and South Asia (e.g., India and Myanmar) exhibit relatively high rural populations (Figure 5(a)), a high proportion of rural population (Figure 5(b)), high values of the employment rate (Figure 5(d)) and the proportion of impoverished population (Figure 5(e)). While the low values of these variables are primarily observed in Europe, North America, and Oceania. Correspondingly, the African and some Asian countries tend to exhibit relatively lower RAI values and higher NSRP values, while the European and some American countries demonstrate relatively higher RAI values and low NSRP values.

Hence, it could be inferred that a higher RAI value signifies the better accessibility of the road system, potentially leading to increase income levels among practitioners and a higher proportion of the population with basic education. And the better road accessibility is associated with the higher level of urbanization, the lower proportion of rural population and decreased poverty levels. Also, a decrease in the NSRP is associated with a increase in both the average income of the residents and the average wage level of the employees.

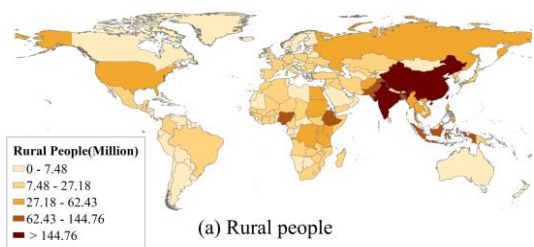
(a) Rural people

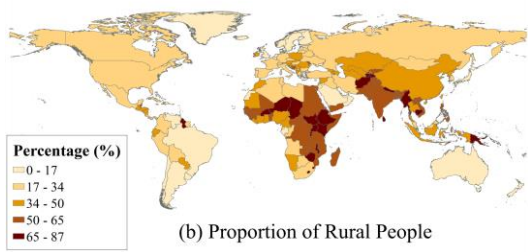
(b) Proportion of Rural People

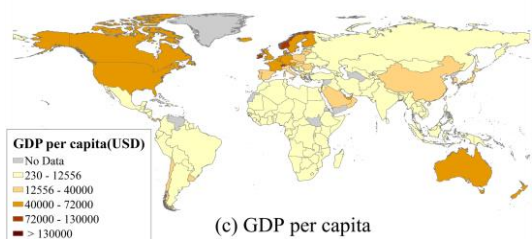
(c) GDP per capita

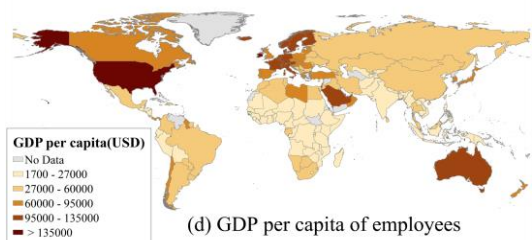
(d) GDP per capita of employees

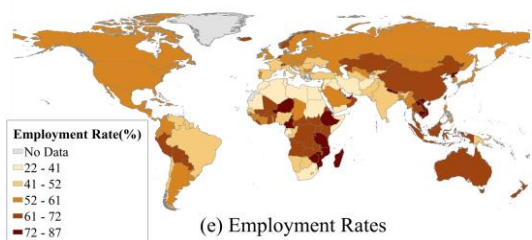
(e) Employment Rates

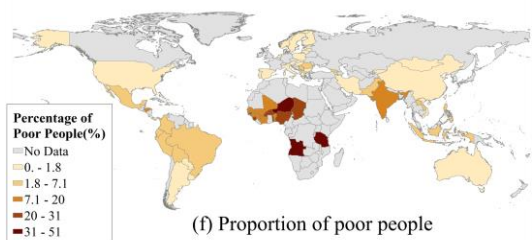
(f) Proportion of poor people

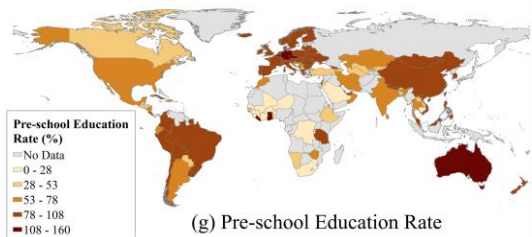
(g) Pre-school Education Rate

Figure 5: Distributions of 7 significant variables

## 5. Conclusions

This study firstly analysed the shortcomings existing in the previous studies and proposed a new indicator, namely the "Not-served rural population (NSRP)", as a complement to RAI for showing the rural accessibility more comprehensively. Then, a new method which is based on the multi-source open data was proposed and applied for calculating the values of RAI and NSRP indicators for 203 countries. Furthermore, this study examined the spatial patterns of these two indicators in space. Lastly, variables in five different dimensions—population, economy, employment, poverty, and education—were chosen to quantitatively explore the relationship between these factors and the spatial patterns of RAI and NSRP. The key conclusions are as follows:

(1) Most countries have RAI values greater than 60%, accounting for 63.04% of the total studying countries. The total NSRP of all these countries is 1.173 billion, indicating that there are so many people who are facing with the challenges of accessing the road services. There is a negative correlation existing between the RAI and NSRP indicators but the correlation coefficient is modest (-0.195). So NSRP is just a complement indicator which could be utilized to measure the number of rural people having difficulties accessing roads more comprehensively. There exists spatial autocorrelation and significant inequality in their global distributions. It is shown that countries with relatively high RAI values (>80%) and low NSRP values (<1 million) are mainly situated in North America, Europe, Eastern Asia, and Oceania, whereas countries in the northern parts of South America, South Asia and Africa generally demonstrate low RAI values (<60%) and high NSRP values (>5 million). More specifically, the countries show significant clustering both in Europe, North America (with high RAI and low NSRP) and Africa (with low RAI and high NSRP), which also reflects that African countries are seriously lagging behind in terms of economic and infrastructure development to some extent.

(2) RAI exhibits a significant positive correlation with per capita GDP, per capita GDP of the employed population, and pre-school education rate. While it shows a significant negative correlation with the proportion of the rural population, the employment rate, and the proportion of the impoverished population. However, The NSRP demonstrates the completely opposite correlations with other relevant factors compared to these relationships of RAI. Since these factors are associated with the level of the development of a country, we can also predict that there also exists a potential positive correlation between this national level of development and RAI. In conclusion, these findings indicate that the economy, population, poverty, and education all have correlations with both RAI and NSRP.

The further explorations of RAI could focus on whether improvements of RAI or NSRP will contribute to the development of socioeconomic, employment, education, etc.